\documentclass[aps,pre,10pt,showpacs]{revtex4-1}

\usepackage{amsmath}
\usepackage{amssymb}
\usepackage{graphicx}
\usepackage{color}

\def\up{{v}}
\def\ul{{u}}
\def\np{{\nu_\perp}}
\def\nl{{\nu_{||}}}
\def\ok{{\overline{\kappa}}}
\def\dn{{\delta n}}
\def\dT{{\delta T}}

\begin{document}

\title{Entropy production in non-equilibrium fluctuating hydrodynamics}

%\date{\today}

\author{Giacomo Gradenigo}
\affiliation{CNR-ISC and Dipartimento di Fisica, Universit\`a Sapienza - p.le A. Moro 2, 00185, Roma, Italy}
\author{Andrea Puglisi}
\affiliation{CNR-ISC and Dipartimento di Fisica, Universit\`a Sapienza - p.le A. Moro 2, 00185, Roma, Italy}
\author{Alessandro Sarracino}
\affiliation{CNR-ISC and Dipartimento di Fisica, Universit\`a Sapienza - p.le A. Moro 2, 00185, Roma, Italy}

\begin{abstract}
Fluctuating entropy production is studied for a set of linearly
coupled complex fields.  The general result is applied to
non-equilibrium fluctuating hydrodynamic equations for coarse-grained
fields (density, temperature and velocity), in the framework of model
granular fluids. We find that the average entropy production, obtained
from the microscopic stochastic description, can be expressed in terms
of macroscopic quantities, in analogy with linear non-equilibrium
thermodynamics. We consider the specific cases of driven granular
fluids with two different kinds of thermostat and the homogeneous
cooling regime. In all cases, the average entropy production turns out
to be the product of a thermodynamic force and a current: the former
depends on the specific energy injection mechanism, the latter takes
always the form of a static correlation between fluctuations of
density and temperature time-derivative. Both vanish in the elastic
limit. The behavior of the entropy production is studied at different
length scales and the qualitative differences arising for the
different granular models are discussed.
\end{abstract}

\pacs{05.40.-a,05.70.Ln,45.70.-n}

\maketitle

%\tableofcontents

%%%%%%%%% INTRO %%%%%%%%%%%%%%%
\section{Introduction}

Among the several different efforts to describe out-of-equilibrium
systems, the linear non-equilibrium thermodynamic approach~\cite{DEGM} has
obtained many important results, at least for a class of systems close
to equilibrium. In this framework, a fundamental role is played by the
macroscopic entropy production, which is related to the irreversible
phenomena occurring in the system due to the presence of heat and
matter currents, chemical reactions, viscous flows, etc...  For
systems in contact with a single thermal bath, the study of
fluctuations relaxing towards the equilibrium state can be described
within such a theory, and extensions of it~\cite{RM00}. Moreover, the
great interest in the study of entropy production in non-equilibrium
stationary states has been motivated by the fact that, in some
particular cases, it can be shown that the stationary state is
characterized by the minimum entropy production principle~\cite{DEGM,P05}.

For generally far from equilibrium systems, a comprehensive theory is
still lacking, but some important results have been obtained in the
last years. In particular, in the framework of systems described by
stochastic models, a stochastic thermodynamics has been proposed,
extending the concepts of standard thermodynamics to fluctuating
quantities, and some general relations have been
proved~\cite{ECM,GC,Kurchan,LS99,CJ,C99,seifert05,HS01,sf12,RXH11}. As
for classical equilibrium thermodynamics, a central role is played by
the microscopic, fluctuating version of entropy production, defined as
functional of finite time-length trajectories in the phase space. Such
a quantity satisfies relations which have been considered by some
authors as generalizations of the second law of
thermodynamics~\cite{Kurchanlast}.

The specific aim of this paper is the study of fluctuations of entropy
production in systems described by a set of coupled Langevin
equations. Many examples of physical systems are well characterized by
such a kind of stochastic process~\cite{R89,sengers,Fr02,HS09}.  In
particular, we are interested in non-equilibrium stationary systems
where the coupling among the different fields defining the model, as
well as the noise entering the Langevin description, violate the
detailed balance condition.  In these out-of-equilibrium systems a
finite rate of entropy production can be measured. However, from a
microscopic point of view, fluctuating entropy production is an
elusive quantity: in a non-equilibrium stationary state (NESS) the
total entropy $S(t)$ (assuming we have a definition for it) has
vanishing time-derivative. In analogy with the perspective of linear
non-equilibrium thermodynamics, one can try to separate this zero
time-derivative into two opposite contributions: one is a surface
term, namely the entropy flow through the boundaries (e.g. heat
exchanged with the thermostat) and the other is a bulk, volume term,
namely the entropy produced inside the system because of internal lack
of detailed balance:
\begin{equation}
\frac{dS}{dt} = \left.\frac{dS}{dt}\right|_{prod}
+\left.\frac{dS}{dt}\right|_{ext}.
\end{equation}
At equilibrium both terms are vanishing, but in a NESS the production
term is always positive. In linear non-equilibrium thermodynamics,
symmetry arguments contribute to make this separation unique; however,
in general, it is not evident that this unique separation can always be
achieved.  A definition of the entropy production term 
related to the dynamics rather than to the thermodynamics of the system has been 
proposed in recent years:
\begin{equation}
\left.\frac{dS}{dt}\right|_{prod} = \lim_{t \to \infty}
\frac{\langle \Sigma(t) \rangle}{t} \ge 0,
\end{equation}
where $\Sigma(t)$ is a functional of the trajectory $\{\phi(s)\}$ of
the system in the time interval $s\in[0,t]$, with $\phi(s)$ a set of
relevant observables. As will be also shown in Sec.~\ref{granular}, 
$\langle \Sigma \rangle$ is often associated
with particular currents (or time-asymmetric correlations) vanishing at
equilibrium. Fluctuations of $\Sigma$ on finite-time trajectories in
finite-size systems, may display also negative values, with an
exponentially small probability, as dictated by the Fluctuation
Relations~\cite{ECM,GC}.

In this paper we approach the problem on fairly general
grounds. Within the framework of Markovian dynamics, we bring to the
fore the structural elements which are common to general systems of
Langevin equations. In particular, we relate the expressions obtained
from the stochastic description of the system to macroscopic
quantities, which can be put in a form very similar to that of
linear non-equilibrium thermodynamics.  

In order to illustrate our general results, we shall focus on the set
of non-equilibrium fluctuating hydrodynamic equations for granular
fluids~\cite{NETP99,GSVP11}. We consider both driven granular gases,
where two different kinds of thermostat are studied, and the case of
homogeneous cooling~\cite{BDKS98}. For such systems, in certain ranges
of the physical parameters, a description in term of coarse-grained
hydrodynamic fields can be given, and a set of coupled Langevin
equations can be written, involving density, temperature and velocity
fields. Applying our formalism to these systems, we are able to
express the average entropy production in terms of macroscopic
currents and thermodynamical forces.  A similar result has been
recently obtained in~\cite{O11}, in the complementary framework of
non-equilibrium stationary spin systems, described by transition rates
violating detailed balance, and in~\cite{sf12}.

The structure of the paper is the following. In Sec.~\ref{irrtherm} we
briefly recall some concepts of linear non-equilibrium thermodynamics. In
Sec.~\ref{fluc} we obtain an expression for the entropy production of
a set of Langevin equations, which is applied to fluctuating
hydrodynamic equations for granular fluids in
Sec.~\ref{granular}. Finally, some conclusions are drawn in
Sec.~\ref{conclusions}. Two Appendices are devoted to some technical
details.

\section{Entropy production in linear non-equilibrium thermodynamics}
\label{irrtherm}

Here we recall some concepts of the linear non-equilibrium thermodynamics
theory~\cite{DEGM}, in order to give a reference frame for the
following discussions.

Linear non-equilibrium thermodynamics is a continuum theory which aims
at describing macroscopic systems characterized by irreversible
processes. The starting point is the balance equation for the entropy,
stating that the entropy in a volume changes because entropy flows
from the boundaries into the volume, or because some irreversible
phenomena are taking place inside the volume.  The underlying
hypothesis is a ``local'' equilibrium assumption, which allows one to
write the thermodynamic Gibbs relations connecting the local (within a
small mass element) entropy with other thermodynamic quantities.

\subsection{Thermodynamic forces and fluxes}
\label{forces}

The first step of the theory is the local formulation of the
conservation laws.  In general, the conservation law for a given
conserved quantity $\rho$ (energy, mass, a component of momentum) in the system reads
\begin{equation}
\frac{\partial \rho(t)}{\partial t}=-\nabla\cdot {\bf J_\rho},
\label{it.1}
\end{equation}
where $J_\rho$ is the flux associated with $\rho$.  For an open
system, in the presence of sources (or sinks) for the quantity $\rho$,
we write the balance equation
\begin{equation}
\frac{\partial \rho(t)}{\partial t}=-\nabla\cdot {\bf J_\rho}+\nu_\rho,
\label{it.1b}
\end{equation}
where $\nu_\rho$ represents the production (or absorption) of quantity
$\rho$ in unit time.

As mentioned above, the variation of entropy $S$ of a macroscopic
system is described by a balance equation, namely it can be split in
two contributions:
\begin{equation}
dS=\left. dS\right|_{ext} + \left. dS\right|_{prod}, 
\label{it.2}
\end{equation}
where $dS|_{ext}$ is the entropy change due to the coupling of the
system with the surrounding medium, and $dS|_{prod}$ is the entropy
produced inside the system due to irreversible processes. For an
insulated system, $dS|_{ext}=0$ and then $dS=dS|_{prod}\ge 0$, from
the second law of thermodynamics. For closed systems exchanging heat
$Q$ with a thermostat at temperature $T$, one has
$dS|_{ext}=\frac{dQ}{T}$ and $dS\ge \frac{dQ}{T}$. The change in time
of the total entropy $S$ is
\begin{equation}
\frac{dS}{dt}=\left. \frac{dS}{dt}\right|_{ext} + \left. \frac{dS}{dt}\right|_{prod}=
-\int_A {\bf J_s}\cdot d{\bf A} + \int_V s~dV,
\label{it.3}
\end{equation}
where $A$ denotes the contour surface of the system, ${\bf J_s}$ is
the entropy flow for unit time and unit area, $s$ is the entropy
production density per unit time and $V$ the total volume.  For open
systems, ${\bf J_s}$ is the sum of the heat flow divided by the local
temperature, ${\bf J_q}/T$, plus all the flows of matters from the
outside. Using the conservation laws and ``local'' equilibrium hypothesis, the entropy
production can be related to the several different irreversible
phenomena occurring inside the system. The structure of the entropy
production $s$ is then that of a bilinear form:
\begin{equation}
s=\sum_i {\bf J_i}\cdot{\bf F_i},
\label{it.4}
\end{equation}
where ${\bf J_i}$ are fluxes (or currents) of the quantities
describing the system, which are associated with irreversible
phenomena, and ${\bf F_i}$ are thermodynamic forces (or affinities),
related to gradients of state variables (spatial non-homogeneities) or
to external forces. In equilibrium conditions, all the thermodynamic
forces vanish and so the entropy production does. Moreover, one also
requires that all fluxes vanish with the thermodynamic forces.

Let us consider for instance the case of a metal bar coupled at
  the edges with two thermostats at different temperatures~\cite{sf12}.  In this
  situation, a stationary temperature profile sets up along the bar,
  with a surface entropy flux due to flux of heat across the edges and
  a bulk entropy production due to the sustain of a temperature
  gradient in the bulk:

\begin{equation}
s=\frac{1}{T}{\bf J_q}\cdot {\bf F_q}=-{\bf J_q}\cdot\frac{\nabla T}{T^2},
\label{it.4a}
\end{equation}
where $T$ is the local temperature and the thermodynamic force
conjugated to the heat flux is ${\bf F_q}=-(\nabla T)/T^2$.  From
  the entropy production in the bulk and the knowledge of the geometry
  of our problem the net flux of entropy across the edges can be then
  deduced. In Sec.~\ref{granular}, we shall consider homogeneously
  driven granular systems in two dimensions, where such a simple
  distinction between surface and bulk contributions cannot be carried
  out.

\subsection{Onsager coefficients}

In order to close the system of equations for the entropy production,
conservation laws and entropy balance equations have to be
supplemented by the phenomenological relations, which, as first
approximation, express the fluxes in terms of a linear combination of
thermodynamic forces
\begin{equation}
J_i=\sum_jL_{ij}F_j,
\label{it.5}
\end{equation}
through the Onsager coefficients $L_{ij}$.  Substituting such
relations into the expression for entropy production, one obtains
\begin{equation}
s=\sum_{i,j}L_{ij}F_iF_j,
\label{it.6}
\end{equation}
namely a quadratic expression in the thermodynamic forces.  In
equilibrium conditions, the thermodynamic forces vanish and, due to
Eq.~(\ref{it.5}), so the fluxes do.

In the physical example introduced above, the phenomenological relation
is the Fourier's law
\begin{equation}
{\bf J_q}=-L_{qq}\frac{\nabla T}{T^2}=-\lambda \nabla T,
\label{it.60a}
\end{equation}
where $\lambda$ is the heat conductivity, and
\begin{equation}
s=L_{qq}\left(\frac{\nabla T}{T^2}\right)^2.
\label{it.60b}
\end{equation}
If the forces $F_i$ perturb a system at equilibrium, i.e. a system
where only conservative (and therefore time-reversible) forces are
present, then Onsager has shown that the coefficients $L_{ij}$ satisfy
the so-called reciprocal relations~\cite{O31,O31b}.

\section{Fluctuating entropy production in stochastic dynamics}
\label{fluc}

Statistical  systems are conveniently described by stochastic
processes.  We recall here how the notion of entropy production can be
introduced in such a framework.  In particular, here we focus on the
degree of irreversibility of the dynamics, as measured by the
functional $\Sigma(\Omega_0^t)$ introduced by Lebowitz and
Spohn~\cite{LS99}, where $\Omega_0^t=\{\phi(s)\}$ is a trajectory of
the phase space in the time interval $s\in[0,t]$. Let us consider for
simplicity a discrete time process, where jumps occur at times $t_i$,
with $i\in[0,n]$.  The functional $\Sigma(\Omega_0^t)$ is defined as
the ratio between the probability of a given trajectory
$P(\Omega_0^t)=p(\phi_0)W(\phi_0|\phi_1)\ldots
W(\phi_{n-1}|\phi_{n})$, where $\phi_i=\phi(t_i)$, and the probability
of time-reversed trajectory $P(\overline{\Omega_0^t})$:
\begin{equation}
\Sigma(t)\equiv\Sigma(\Omega_0^t)
=\ln\frac{P(\Omega_0^t)}{P(\overline{\Omega_0^t})}-\ln\frac{p(\phi_0)}{p(\phi_t)}=\ln
\frac{W(\phi_0|\phi_1)\ldots
  W(\phi_{n-1}|\phi_{n})}{W(\phi_n|\phi_{n-1})\ldots
  W(\phi_1|\phi_0)},
\label{it.10b}
\end{equation}
where $\overline{\Omega_0^t}\equiv \{\overline{\phi}(t-s)\}$ and
$\overline{\phi}=\epsilon \phi$, with $\epsilon=\pm 1$, according to
the parity of the field.  

In a stationary state, where entropy produced/consumed inside
the system is continuously balanced by a flux coming from the
boundaries, as in Eq.~(\ref{it.3})
\begin{equation}
0=\frac{dS}{dt}=\left. \frac{dS}{dt}\right|_{ext} + \left. \frac{dS}{dt}\right|_{prod},
\label{it.11}
\end{equation}
we write, following~\cite{LS99},

\begin{equation}
\left.\frac{dS}{dt}\right|_{prod} = \lim_{t \to
  \infty}\frac{1}{t}\left\langle \Sigma(t) \right\rangle \ge 0.
\label{it.12}
\end{equation}
The functional $\Sigma(t)$ depends on a stochastic variable (i.e. a
trajectory of a stochastic differential equation) and is itself a
stochastic variable with a given probability density: indeed, for
finite $t$, it can also take negative values, of course with a
probability decreasing with increasing $t$. At large times and for
bounded (or negligible) values of the term $\ln \frac{p(\phi_0)}{p(\phi_t)}$, the
probability density of $\Sigma$ satisfies the so-called
Fluctuation Relation~\cite{ECM,GC,LS99,seifert05,ZC03,PRV06}.

In the following we will show that, in cases which are relevant for
non-equilibrium spatially extended systems, the
functional~(\ref{it.10b}) can take a clear thermodynamic meaning,
since it can be expressed in terms of macroscopic currents and forces,
in strict analogy with the form of entropy production~(\ref{it.4})
discussed in Sec.~\ref{forces}.

\subsection{Coupled Langevin equations}

Here we specialize to a continuous space and time Markov process and,
hereafter, $\phi$ denotes a complex $d_\phi$-dimensional vector, with
components $\phi_i$ ($i=1,\ldots,d_\phi$).  The time-reversal
transformation on vector $\phi$ is defined as an operator which
changes $\phi_i \to \overline{\phi_i}\equiv \epsilon_i \phi_i$ with
$\epsilon_i \in \{+1,-1\}$: this implies that both real and imaginary
parts of $\phi_i$ have the same sign-change upon time-reversal. As
shorthand notations we will use $\overline{\phi}$ or $\epsilon \phi$
to indicate the vector made of time-reversed components $\{\epsilon_i
\phi_i\}$.  In a coarse-grained description of a fluid system,
  instances of even variables are the Fourier components of local
  density and temperature fluctuations ($\epsilon_i=+1$), whereas the
  velocity field is odd ($\epsilon_i=-1$). We assume that the dynamics
  of each component of the vector $\phi$ is described by a Langevin
  equation:

\begin{equation} \label{fw_eq}
\frac{d\phi_i(t)}{dt}=B_i[\phi(t)]+\xi_i(t),
\end{equation}
with $B_i$ a complex drift and $\xi_i$ a Gaussian process with
$\langle \xi_i(t)\xi^*_j(s)\rangle=2\delta_{ij}\delta(t-s)D_{ii}$,
where the $x^*$ denotes the complex conjugate of $x$. If we consider a
trajectory between time $0$ and time $t$, we can define its
time-reversal as $\overline{\phi_i(s)}=\epsilon_i\phi_i(t-s)$.  For
the complementary discussion starting from the corresponding
Fokker-Planck equation, see Appendix~\ref{app2}.  The stochastic
  system in Eq.~(\ref{fw_eq}) is indeed a good model to describe the
  behavior of macroscopic variables in many contexts~\cite{sengers}.
  Often it works as a first approximation, where one retains only the
  linear part of the dynamics and the effect of nonlinearities is
  replaced with noise terms.

 In order to define a useful projection of the dynamics, let us for
  the moment neglect the noise terms in Eq.~(\ref{fw_eq}), namely
  $\xi_i \equiv 0$.  The time-reversal trajectory satisfies the
following differential equation:
\begin{equation} 
\frac{d\overline{\phi_i(s)}}{ds}=-\epsilon_i
\frac{d\phi_i(t-s)}{dt}=-\epsilon_i B_i[\phi(t-s)]=-\epsilon_i
B_i[\epsilon\overline{\phi(s)}].
\end{equation}
We notice two particular cases for $B_i(\phi)$:
\begin{align}
B_i(\epsilon\phi)&=-\epsilon_i
B_i(\phi)\\ B_i(\epsilon\phi)&=\epsilon_i B_i(\phi).
\end{align}
In the first case $\overline{\phi_i(s)}$ satisfies exactly the forward
equation~(\ref{fw_eq}). In the second case it satisfies the same
equation with drift changed of sign.  Following these two limit cases,
we will in general decompose the $i$-th component of the drift
as~\cite{R89}
\begin{equation}
B_i(\phi) = B_{i,rev}(\phi)+B_{i,irr}(\phi),
\end{equation}
where
\begin{align}\label{rev}
B_{i,rev}(\phi)&=\frac{1}{2}[B_i(\phi)-\epsilon_iB_i(\epsilon \phi)]=-\epsilon_iB_{i,rev}(\epsilon\phi)\\
B_{i,irr}(\phi)&=\frac{1}{2}[B_i(\phi)+\epsilon_iB_i(\epsilon \phi)]=\epsilon_iB_{i,irr}(\epsilon\phi),
\label{irr}
\end{align}
where $B_{i,rev}$ and $B_{i,irr}$ represent the reversible and
irreversible contributions to the drift, respectively.

%%%%%%%%%%%%%%%%%%%%%%%%%%%%%%%%%%%%%%%%%%%%%%%%%%%%%%%%%%%%%%%%%
\subsection{Entropy production functional}

 We assume that the system~(\ref{fw_eq}) converges to a unique
  stationary state in the long time limit.  Then, let us consider the
  entropy production in such a stationary state: following the
  Onsager-Machlup prescription, in the case of an additive Gaussian
  noise we can write the probability of a trajectory
  $\Omega_0^t\equiv\{\phi(s),s=0\cdots t\}$ as:
\begin{equation}
P(\Omega_0^t)=p[\phi(0)]\exp[\mathcal{S}(\Omega_0^t)], \qquad
P(\overline{\Omega_0^t})=p[\overline{\phi(0)}]\exp[\mathcal{S}(\overline{\Omega_0^t})],
\end{equation}
where $p(\phi)$ is the weight (probability or density) of state $\phi$
in the steady state, and the {\em action} $\mathcal{S}$ is given
by~\cite{OM53}
\begin{eqnarray}
\mathcal{S}(\Omega_0^t)&=&-\int_0^t ds
\sum'_i\left\{\frac{1}{4D_{ii,R}}\left[\frac{d\phi_i(s)}{ds}-B_i[\phi(s)]\right]
\left[\frac{d\phi^*_i(s)}{ds}-B^*_i[\phi(s)]\right]\right\}.
\end{eqnarray}
We have used the notation $\sum'_i$ to indicate that the sum runs only
on those indexes such that $D_{ii}\neq 0$~\cite{MO53}.  Due to
  stationarity, we can always set to zero the initial time of the
  interval where the entropy production is calculated.  Recalling
that $\int_0^t ds f(t-s)=\int_0^t ds f(s)$, we easily get for the
entropy produced by a path $\Omega_0^t$~\cite{PV09}:
\begin{equation}
\Sigma(\Omega_0^t) = \ln
\frac{P(\Omega_0^t)}{P(\overline{\Omega_0^t})}-
\ln\frac{p[\phi(0)]}{p[\overline{\phi(t)}]}= \mathcal{S}(\Omega_0^t)
-\mathcal{S}(\overline{\Omega}_0^t)=\int_0^t ds~\sigma(s),
\label{definition}
\end{equation}
with entropy production rate
\begin{equation}
\sigma(s)=\sum'_i\frac{1}{2D_{ii}}\underbrace{B_{i,irr}[\phi(s)]}_{force}
\underbrace{(\dot{\phi}^*_i(s)-B^*_{i,rev}[\phi(s)])}_{current}+c.c.,
\end{equation}
where c.c. denotes the complex conjugate.  In this formalism, we can
point out how the entropy production is expressed as the product of
forces (the drift) multiplied by currents (time derivative of fields).
 According to the phenomenological relations~(\ref{it.5}), the
  irreversible fluxes vanish when the driving forces are switched
  off. Our identification of forces and fluxes is consistent with such
  relations: indeed, when $\langle B_{i,irr}[\phi(s)]\rangle
  =0$ we also have
  $\langle\dot{\phi}^*_i(s)-B^*_{i,rev}[\phi(s)]\rangle = 0$, as can
  be seen by averaging equation~(\ref{fw_eq}) over the noise.  This
is in analogy with the macroscopic expressions in linear
non-equilibrium thermodynamics.

%%%%%%%%%%%%%%%%%%%%%%%%%%%%%%%%%%%%%%%%%%%%%%%%%%%%%%%%%%%%%5
\subsection{Linear processes}

We consider now the linear case $B[\phi]\equiv A\phi$ where $A$ is the
so-called dynamical matrix:
\begin{equation} \label{lin_eq}
B_i[\phi]=\sum_j A_{ij}\phi_j=\sum_{j\in \text{rev}(i)}A_{ij}\phi_j+\sum_{j\in \text{irr}(i)}A_{ij}\phi_j,
\end{equation}
where the shorthand notation $j\in\text{rev}(i)$ means that the index
$j$ runs on the set of indices such that $\epsilon_j=-\epsilon_i$,
while $j\in\text{irr}(i)$ stands for the set of indices such that
$\epsilon_j=\epsilon_i$.  In this case one has

\begin{eqnarray} \label{linear_ep}
\sigma &=& \sum'_i\frac{1}{2D_{ii}}\underbrace{\left(\sum_{j\in
    \text{irr}(i)}A_{ij}\phi_j\right)}_{force}\underbrace{\left[\dot{\phi}^*_i-\sum_{j\in
      \text{rev}(i)}A^*_{ij}\phi^*_j\right]}_{current}+c.c. \\ 
&=& \sum'_i\frac{1}{2D_{ii}}\left[\sum_{j\in
    \text{irr}(i)}A_{ij}\left(\phi_j\dot{\phi}^*_i+c.c.\right)
  +\sum_{\substack{j\in \text{irr}(i)\\l\in
      \text{rev}(i)}}A_{ij}A_{il}\left(\phi_j\phi^*_l-c.c.\right)\right].
\label{linear_ep2}
\end{eqnarray}
In Appendix~\ref{app1} we show that at equilibrium $\langle\sigma
\rangle=0$, as expected.  Formula~(\ref{linear_ep}) is analogous to
the macroscopic thermodynamic result~(\ref{it.4}): it expresses the
entropy production as the product of a force by a current. However, in
the general case, the formula remains rather abstract, and we need
explicit examples to illustrate its meaning.  These will be discussed
in the following section.

%%%%%%%%%%%%%%%%%%%%%%%%%%%%%%%%%%%%%%%%%%%%%%%%%%%%%%%%%%%%%%
\section{Applications to granular fluids}
\label{granular}

In this section we show three examples in the framework of
granular fluids where, starting from the definition~(\ref{definition}),
we can obtain an expression for the entropy production similar
to Eq.~(\ref{it.4}). 

\subsection{Fluctuating hydrodynamics for driven granular fluids}

We consider a fluid of $N$ identical inelastic hard spheres in
dimension $d$, of mass $1$ and diameter $r$, in a square box of volume
$V$ with external homogeneous stochastic driving. We denote by $\rho$
the packing fraction of the system, which for $d=2$ reads
$\rho=N\pi(r/2)^2/V$.  The model is defined by giving the equation of
motion for the velocity of $i$-th particle:
\begin{equation}
\dot{v}_i(t)=-\gamma_b v_i(t) +\sqrt{2T_b\gamma_b}\zeta_i(t)+F_i(t),
\label{micro1}
\end{equation}
where $\gamma_b$ is a viscous drag, $T_b$ is the temperature of the
external thermostat, $\zeta_i$ is a Gaussian noise with $\langle
\zeta_i \rangle=0$ and $\langle \zeta_i(t) \zeta_j(t')
\rangle=\delta_{ij}\delta(t-t')$ and $F_i(t)$ is the
resulting force of eventual instantaneous collisions with other
particles. Every time two particles $i$ and $j$ collide, their
velocities are instantaneously changed following the rule
\begin{equation}
v_i'=v_i-\frac{1+\alpha}{2}[(v_i-v_j)\cdot \hat{n}]\hat{n},
\label{collrule}
\end{equation}
where $\alpha\in[0,1]$ is the restitution coefficient ($=1$ for
elastic collisions) and $\hat{n}$ is the unit vector in the direction
joining the centers of the colliding particles.  This model has been
introduced in~\cite{PLMPV98} and, recently, it has been also
  successfully used to describe real granular
  experiments~\cite{GSVP11bis,PGGSV12}. A different model thermostat
is obtained in the limit $\gamma_b \to 0$, $T_b \to \infty$ with
constant $\Gamma=2\gamma_b T_b$, as it has been studied
in~\cite{NETP99} (see Sec.~\ref{trizac}). In both cases the fluid
reaches a NESS with a well defined granular temperature $T=\langle
|v|^2 \rangle/d$. In the case of finite $\gamma_b>0$, a characteristic
thermostat time $\tau_b=1/\gamma_b$ is defined. When compared to the
mean collision time $\tau_c=1/\omega$ (with $\omega$ the collision
frequency), it determines two possible regimes: $\tau_b\gg\tau_c$ is
the dissipative regime, where $T \le T_b$ (the equal sign holding only
for $\alpha=1$); $\tau_b\ll\tau_c$ is the equilibrium regime where
$T=T_b$ for any $\alpha$, since collisions are negligible.  

Let us add a brief comment on such a model, slightly anticipating
  some considerations on the entropy production to be discussed below.
  From Eq.~(\ref{micro1}), we see that the thermostat is homogeneously
  coupled to \emph{all} the particles. Therefore, the energy injection
  represents by definition a bulk contribution. It is then through
  collisions, Eq.~(\ref{collrule}), that the dissipation of energy
  takes place. Because collisions occur across the whole system, also
  the energy sink represent a bulk contribution. Therefore, we can imagine that
  inelastic collisions are perfectly equivalent to a zero temperature
  reservoir, which is coupled to all the particles. In such a description,
  what is generally called ``entropy flux'' refers to the
  \emph{entropy exchanged with reservoirs}, rather than to entropy
  flowing into the system through the boundaries.

In the case where a separation of temporal and spatial scales takes
place, a hydrodynamic description of the system can be given. Then, we
introduce the coarse-grained hydrodynamic fields, density, velocity
and temperature, $n({\bf r},t),{\bf u}({\bf r},t)$ and $T({\bf r},t)$,
respectively, where ${\bf r}$ denotes a point in the $d$-dimensional
space, as follows:
\begin{eqnarray}
n({\bf r},t)&=&\sum_i\delta({\bf r}-{\bf r}_i(t)), \nonumber \\
{\bf u}({\bf r},t)&=&\frac{1}{n}\sum_i {\bf v}_i(t)\delta({\bf r}-{\bf r}_i(t)), \label{hf} \\
T({\bf r},t)&=&\frac{2}{dn}\sum_i \frac{v^2_i(t)}{2}\delta({\bf r}-{\bf r}_i(t)). \nonumber 
\end{eqnarray}
Here the sums are over all the particles in the system.  The
homogeneous stationary state is characterized by constant density $n$
and granular temperature $T$. The fluctuations of hydrodynamic fields
are $\{\delta n=n(r,t)-n, \delta T=T(r,t)-T, u_x,u_y\}$ and are
described by the set of linear hydrodynamic equations.  In these
equations the small scale fluctuations have been projected out, but
their effect on large scale fluctuations can be taken into account by
a proper addition of noise terms~\cite{sengers}, resulting in a
fluctuating hydrodynamic description.  For these models, such
equations have been studied in~\cite{NETP99,GSVP11}.

 After
  linearization near the homogeneous state, and changing to Fourier
  space, in two spatial dimensions ($d=2$), we are left with a linear
  Langevin system for the components of a four-dimensional
  ($d_\phi=4$) complex field vector
  $\phi=(\dn(k),\dT(k),\ul(k),\up(k))$, with parities under
  time-reversal $\epsilon=(1,1,-1,-1)$. By $\ul$ and $\up$ we mean
the longitudinal and transverse velocity field, respectively. In this
case a useful shorthand notation consists in replacing the index $i$
by a label equal to the name of the fields (omitting the $\delta$s for
simplicity), i.e. $i=1 \to n$, $i=2 \to T$, $i=3 \to u$ and $i=4 \to
v$. The Langevin equation for each component of the vector field is
then

\begin{equation}\label{langgran}
\dot{\phi_i}=A_{ij}\phi_j+\xi_i,
\end{equation}
where the specific form of the dynamic matrix $A$ and of the noise
amplitudes depend on the kind of thermostat and will be explicitly
given below.  In this case, from the definitions following
Eq.~(\ref{lin_eq}) we have

\begin{align}
\text{irr}(n)&=(n,T)\;\;\;\;&\text{rev}(n)=(\ul,\up)\\
\text{irr}(T)&=(n,T)\;\;\;\;&\text{rev}(T)=(\ul,\up)\\
\text{irr}(\ul)&=(\ul,\up)\;\;\;\;&\text{rev}(\ul)=(n,T)\\
\text{irr}(\up)&=(\ul,\up)\;\;\;\;&\text{rev}(\up)=(n,T),
\end{align}
and the set of irreversible and reversible coefficients of the matrix
are those with indexes given by, respectively:

\begin{align}
\text{irr}&=(nn,nT,Tn,TT,\ul\up,\ul\ul,\up\ul,\up\up)\\
\text{rev}&=(n\ul,n\up,T\ul,T\up,\ul n,\ul T,\up n,\up T).
\end{align}
In the following we consider the two kinds of thermostat introduced
above.

\subsubsection{Finite temperature thermostat, $\gamma_b>0$}

In the case $\gamma_b>0$, the dynamic matrix reads

\begin{equation}A=
\begin{pmatrix}
  0                                &0                                    &0                      &0\\
-\gamma_0\frac{\omega g_2 T}{n}   &-(3\gamma_0\omega+\ok k^2+2\gamma_b)  &0                      &0\\
 0                                 &0                                    &-(\nl k^2+\gamma_b)    &0\\
 0                                 &0                                    &0                      &-(\np k^2+\gamma_b)
\end{pmatrix}\\+\mathcal{I}
\begin{pmatrix}
  0                                &0                                    &-k n                      &0\\
  0                                &0                                    &-k \frac{2p_0}{nd}           &0\\
 -k \frac{c^2}{n}                &-k\frac{p_0}{mnT}                  &0                           &0\\
 0                                 &0                                    &0                          &0
\end{pmatrix}.
\label{A}
\end{equation}
where $\np$ and $\nl$ are the kinematic shear and longitudinal
viscosity respectively, while $\ok=2\kappa/(nd)$ is the thermal
diffusion coefficient associated with the thermal conductivity
$\kappa$.  Here $\gamma_0=(1-\alpha^2)/2d$, $g_2$ is the pair
correlation function at contact (in two dimensions we use the
Verlet-Levesque approximation $g_2=(1-7\rho/16)/(1-\rho)^2$, $p_0$ is
the pressure, which for elastic hard disks reads $p_0=nT(1+2\pi
g_2nr^2/4)$, and $c$ is the thermal velocity. Here and in the
following the transport coefficients $\np$, $\nl$ and $\ok$ are
evaluated by the Enskog theory for elastic disks~\cite{CC70} at
temperature T (which is the NESS temperature and may depend on
inelasticity). Note that irreversible coefficients of $A$ are always
real, while reversible coefficients are always imaginary.

Hydrodynamic noise is the sum of external noise due to the bath and
internal one due to collisions. Only the latter is assumed to satisfy
a ``local equilibrium'' assumption, i.e. the fluctuation-dissipation
relation with the granular temperature~\cite{NETP99,GSVP11}. Noises
amplitudes have, then, the following amplitudes:
\begin{eqnarray}
D_{nn}&=&0\\
D_{TT}&=&\frac{4 \gamma_b T_b T}{d n} + \frac{2k^2 \ok T^2}{n d}\\
D_{\ul\ul}&=&\frac{\gamma_b T_b}{n} + \frac{\nl k^2 T}{n}\\
D_{\up\up}&=&\frac{\gamma_b T_b}{n} + \frac{\np k^2 T}{n}.\label{dvv}
\end{eqnarray}

Following the general formula~\eqref{linear_ep2}, we get - for the entropy production rate at a given $k$:

\begin{equation}
\sigma_k=\left[\frac{A_{Tn}}{D_{TT}}+\frac{i}{kn}\frac{A_{TT}A_{Tu}}{D_{TT}}-\frac{i}{kn}\frac{A_{uu}A_{uT}}{D_{uu}} \right]
\frac{\dn\dot{T}^*+c.c.}{2}+b.t.
\label{general}
\end{equation}
To obtain the above formula we have used the linearized continuity
equation $\dot{\dn}=-i(kn)u$ which allows one to do the following
replacements: $u=i\dot{\dn}/kn$ and $u^*=-i\dot{\dn}^*/kn$. Moreover,
we have exploited the fact that all terms of the kind
$\phi\dot{\phi}^*+c.c.$ are so-called ``boundary terms'', i.e. they
contribute to the entropy produced in the time-interval $[0,t]$ by an
amount $|\phi(t)|^2-|\phi(0)|^2$, which is {\em usually} sub-leading
with respect to other terms which grow with time
$t$~\cite{ZC03,PRV06,germans,BGGZ06}. We have denoted all these terms
by the symbol $b.t.$ (those terms vanish in long-time averages, as
dicussed below). Using the definition of the dynamical matrix $A$ in
Eq.~(\ref{general}), we have

\begin{equation}
\sigma_k=h(k)\Re[\dn(k)\dot{T}^*(k)]+b.t.,
\label{nT}
\end{equation}
where 

\begin{equation}
h(k)=-\frac{(T_b-T)}{nT}\frac{\gamma_b\left\{p_0[\ok +2(3/d-1)\nl]k^2
+g_2n(\gamma_bT_b+\nl k^2T)+6p_0\gamma_bT_b/(dT)\right\}}{(\ok k^2T+2\gamma_bT_b)(\nl k^2T+\gamma_bT_b)}.
\end{equation}
Here $\Re[x]$ denotes the real part of $x$ and we have used the
relation $\gamma_0\omega\simeq 2\gamma_b(T_b-T)/(dT)$~\cite{SVCP10}. At
equilibrium, namely for elastic interactions, $T=T_b$ and the
entropy production vanishes identically, as expected.

Notice that in the linearized hydrodynamics, modes at different $k$ do
not interact and therefore do not exchange energy or produce
entropy. Entropy is instead produced separately at each $k$, because
of unbalanced fluxes among different components of the fields, i.e. an
entropy production rate $\sigma_k$ exists for each mode $k$.  The
total rate of entropy produced by the hydrodynamic degrees of freedom
is the sum of all $\sigma_k$ in the hydrodynamic range of $k$s.
Interestingly enough, because of the linear approximation adopted,
each scale $k$ has its own entropy production which is expected to
satisfy the Fluctuation Relation.

\subsubsection{Average entropy production}

\begin{figure}[!t]
\begin{center}
\includegraphics[width=0.6\columnwidth,clip=true]{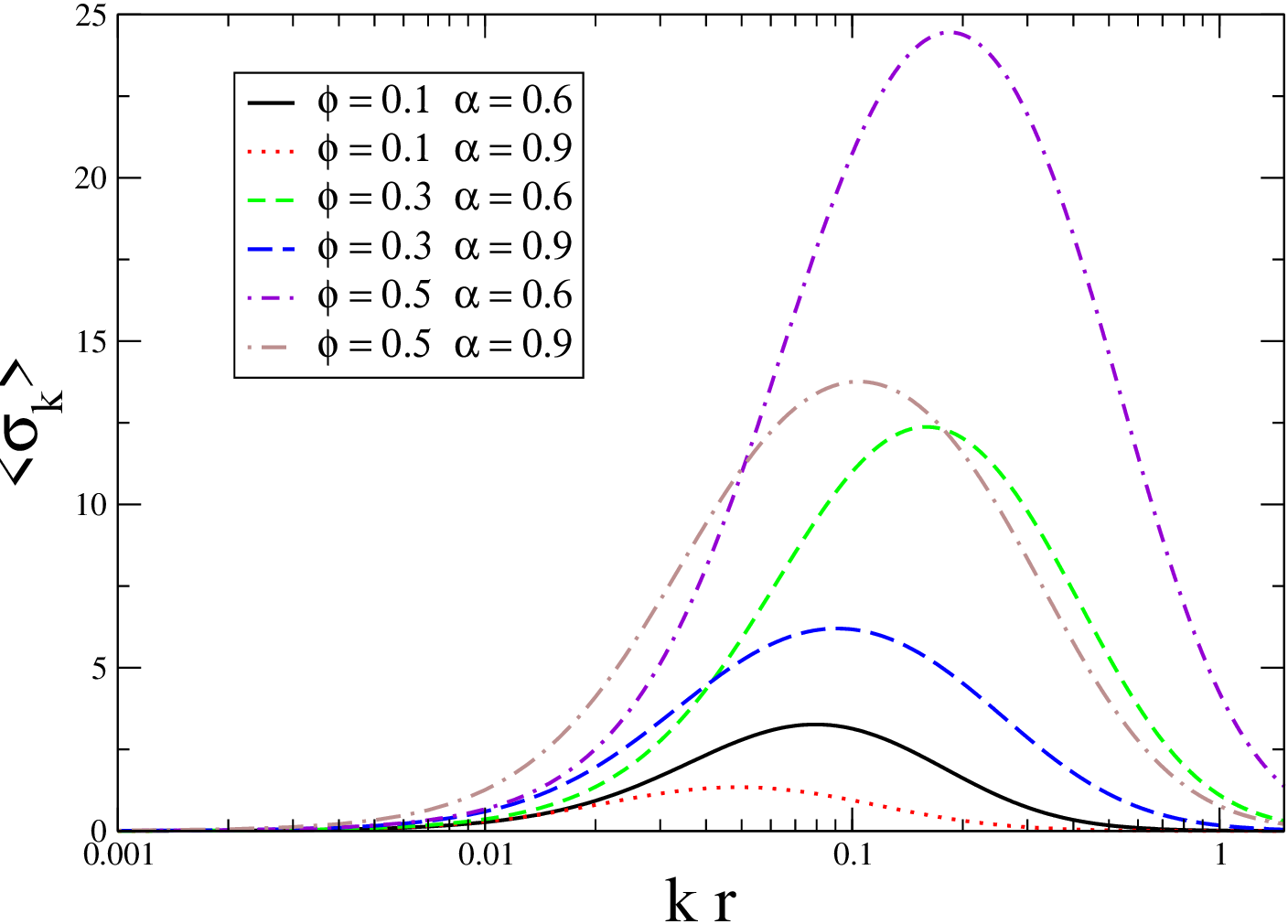}
\caption{Average entropy production $\langle \sigma_k\rangle$ as a
  function of $kr$, Eq.~\eqref{nT}, for the driven granular model with finite
  temperature thermostat and friction $\gamma_b>0$ (defined by the
  Eqs.~(\ref{langgran}) and~(\ref{A}-\ref{dvv})). The different curves
  show the behavior of $\langle \sigma_k\rangle$ for several values of
  the parameters $\alpha$ and $\rho$, as reported in the
  legend. Notice the peak at a certain wave-vector which signals the
  presence of a characteristic length-scale, as explained in the
  text.}
\label{noi}
\end{center}
\end{figure}

Eq.~(\ref{nT}) is the central result of this paper: it expresses the
entropy production defined microscopically in terms of macroscopic
quantities, following the structure of linear non-equilibrium thermodynamics. In
particular, the quantity $\delta n(k)\dot{T}^*(k)$ plays the role of a current of
energy, whereas the quantity $h(k)$ is a thermodynamic
force which is vanishing at equilibrium.  Moreover, taking the
stationary average over the stationary distribution on
equation~(\ref{nT}), the boundary terms vanish, and, exploiting the
dynamical equation for the temperature field, the entropy production
can be expressed as a linear combination of static correlation
functions (structure factors)

\begin{equation}
\langle\sigma_k\rangle=h(k)\Re[\langle\dn(k)\dot{T}^*(k)\rangle]
=h(k)[A^*_{Tn}C_{nn}(k)+A^*_{TT}C_{nT}(k)+A^*_{Tu}C_{nu}(k)],
\label{nT2}
\end{equation}
where $C_{ij}(k)=\langle \phi_i(k)\phi_j^*(k)\rangle$. The structure
factors can be computed analytically. Then, for small $k$, we have
$h(k)\sim a_0+a_2 k^2$, with

\begin{equation}
a_0=-\frac{(6 p_0 + d g_2 n T) \gamma_0 \omega_0}{4n T T_b \gamma_b}
\end{equation}
and $\Re[\langle\dn(k)\dot{T}^*(k)\rangle] \sim b_2 k^2$, so that

\begin{equation}
\langle \sigma_{k=0}\rangle=0.
\end{equation}
In the opposite limit, $k\to\infty$, one finds the behavior
$\langle\sigma_k\rangle\sim k^{-4}$.  In Fig.~\ref{noi} we show the
average entropy production as a function of $k$ for several different
values of the parameters $\alpha$ and $\rho$. The first observation is
that the average entropy production defined in Eq.~(\ref{nT2})
vanishes by definition at equilibrium.  In this situation, the
coefficient $A^*_{Tn}$ is zero and the correlation functions
$C_{nT}(k)$ and $C_{nu}$ between fields of opposite parity under time
reversal, vanish due to time reversibility. Moreover, from the results
of our calculations we find that, even in {\em inelastic cases},
both at very large and very short length scales, the system looks like
in equilibrium, where we have a zero entropy production. From the
results of a previous work~\cite{GSVP11}, we know that modes at $k
\rightarrow 0$ show equipartition at the bath temperature $T_b$,
whereas modes for $k$ very large at the granular temperature $T$.  It
makes sense that at the length-scales where a good degree of
equipartition is attained the entropy production turns to be
zero. Another interesting similarity with the results on fluctuating
hydrodynamics obtained in~\cite{GSVP11} is that, also here, from the
study of the entropy production, a characteristic
\emph{non-equilibrium} length-scale emerges. It seems that there is a
preferential wave-vector at which the excitation of modes in the
system produces the higher degree of dissipation, and that this
wave-vector corresponds to a finite length.

\subsubsection{Infinite temperature thermostat}
\label{trizac}

For the case $\gamma_b=0$ and $T_b \to \infty$ with finite $\Gamma=2 \gamma_b T_b$, the dynamical matrix reads

\begin{equation}A=
\begin{pmatrix}
  0                                &0                                    &0                      &0\\
-\gamma_0\frac{\omega g_2 T}{n}   &-(3\gamma_0\omega+\ok k^2)             &0                      &0\\
 0                                 &0                                    &-\nl k^2               &0\\
 0                                 &0                                    &0                      &-\np k^2
\end{pmatrix}\\+\mathcal{I}
\begin{pmatrix}
  0                                &0                                    &-k n                      &0\\
  0                                &0                                    &-k \frac{2p_0}{nd}           &0\\
 -k \frac{c^2}{n}                &-k\frac{p_0}{mnT}                  &0                           &0\\
 0                                 &0                                    &0                          &0
\end{pmatrix},
\label{Atriz}
\end{equation}
and noise amplitudes are
\begin{eqnarray}
D_{nn}&=&0\\
D_{TT}&=& \frac{1}{2}\left(\frac{4 T\Gamma}{d n} + \frac{4k^2 \ok T^2}{n d}\right)\\
D_{\ul\ul}&=&\frac{1}{2}\left(\frac{\Gamma}{n} + \frac{2\nl k^2 T}{n}\right)\\
D_{\up\up}&=&\frac{1}{2}\left(\frac{\Gamma}{n} + \frac{2\np k^2 T}{n}\right).\label{dvvtrizac}
\end{eqnarray}
Then, following exactly the same computation as before, we find

\begin{equation} \label{nTinf}
\langle\sigma_k\rangle=h'(k) \Re[\langle\dn(k)\dot{T}^*(k)\rangle]
\end{equation}
with

\begin{equation}
h'(k)=-\frac{\gamma_0\omega_0\{[2p_0(\ok +\nl)+dg_2 nT\nl]k^2+(6p_0+dg_2nT)\gamma_0\omega_0\}}
{2nT^2_g(\ok k^2+2\gamma_0\omega_0)(\nl k^2+\gamma_0\omega_0)}. 
\end{equation}

\begin{figure}[!t]
\begin{center}
\includegraphics[width=0.6\columnwidth,clip=true]{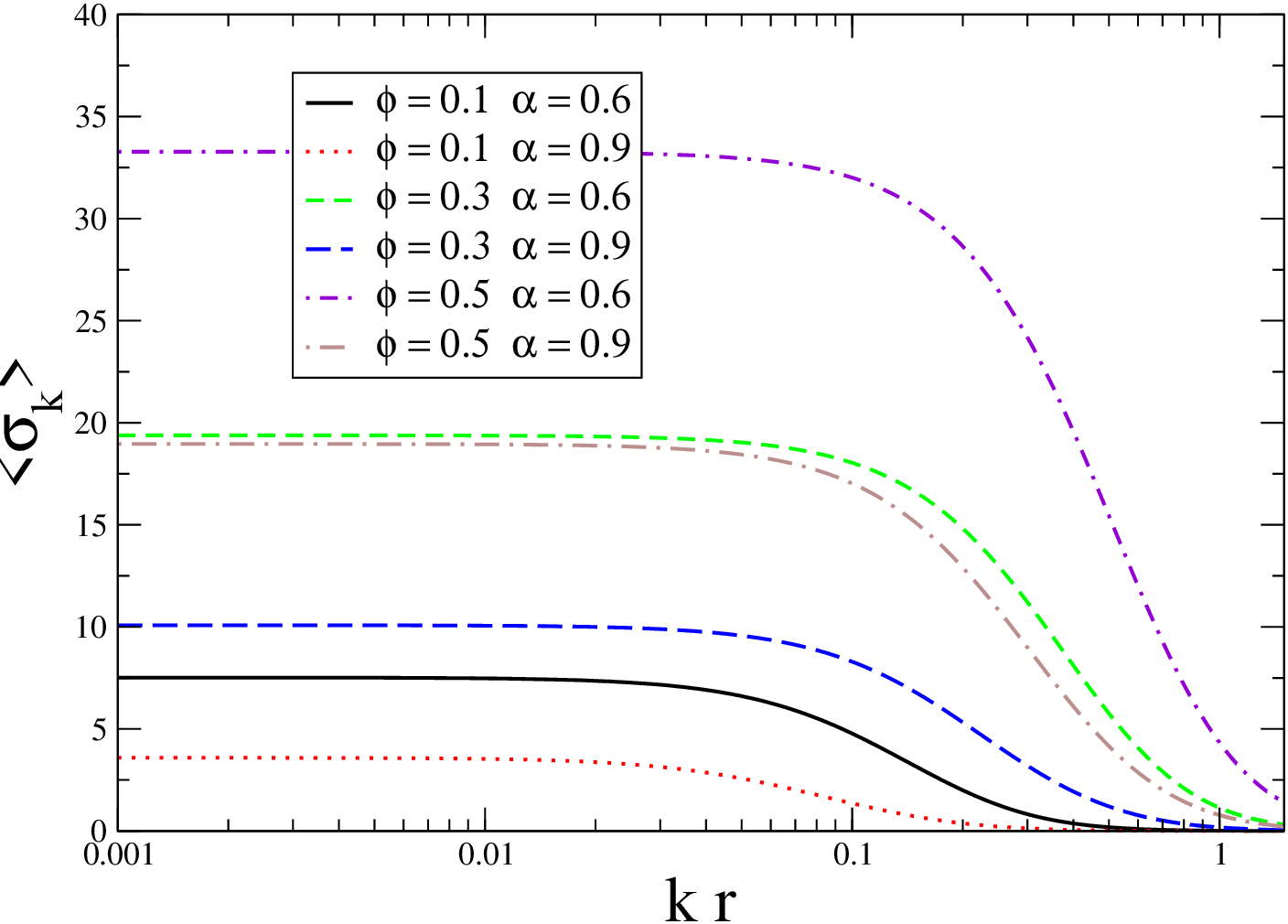}
\caption{Average entropy production $\langle \sigma_k\rangle$ as a
  function of $kr$, Eq.~\eqref{nTinf}, for the driven granular model with infinite
  temperature thermostat and without friction, $T_b=\infty$,
  $\gamma_b=0$ with $2\gamma_b T_b=\Gamma$ finite, defined by the
  Eqs.~(\ref{langgran}) and~(\ref{Atriz}-\ref{dvvtrizac}). The
  different curves show the behavior of $\langle \sigma_k\rangle$ for
  several values of the parameters $\alpha$ and $\rho$, as reported in
  the legend. Notice that $\sigma_k$ reaches a finite value for $k\to
  0$.}
\label{trizacf}
\end{center}
\end{figure}

In this case, the series expansion of
the prefactor $h'(k)$, for small $k$, gives $h'(k)\sim
a'_0+a'_2 k^2$, with

\begin{equation}
a'_0=-\frac{6 p_0 + d g_2 n T}{4n T^2},
\end{equation}
while $\langle \Re[\dn(k)\dot{T}^*(k)]\rangle \sim b_0'+b_2' k^2$, with

\begin{equation}
b_0'=-\frac{n T^2 (6 p_0 + d g_2 n T) \gamma_0 \omega_0}{ 6 p_0^2 + d
  g_2 n p_0 T + 9 d n^2 T \gamma_0 \nl \omega_0},
\end{equation}
so that for the entropy production at $k=0$ we have the finite value

\begin{equation}
\langle \sigma_{k=0}\rangle=\frac{(6 p_0 + d g_2 n T)^2 \gamma_0 \omega_0}{4
  (6 p_0^2 + d g_2 n p_0 T + 9 d n^2 T \gamma_0 \nl \omega_0)}.
\end{equation}
This result is qualitatively different from that obtained in the case
of a finite temperature thermostat.  The finite value for $k\to 0$
obtained in this case could be related to the scale-free correlations
in the fluctuations of the hydrodynamic fields of this
model~\cite{TPNE01}. In the opposite limit, $k\to\infty$, we again
find $\langle \sigma_k\rangle\sim k^{-4}$. In Fig.~\ref{trizacf} we
show the average entropy production as a function of $k$ for several
different values of the parameters $\alpha$ and $\rho$. Notice here
that there is always a finite entropy production, even at very short
wave-vectors, where a good degree of equipartition between modes is
never obtained.  The result we find is consistent with the fact that
it does not exist any finite length-scale at which we can coarse grain
the local hydrodynamics fields in order to see an equilibrium-like
behavior.

\subsection{Homogeneous cooling}

In the limit $\Gamma\to 0$, the above model becomes a pure cooling
granular fluid (no energy injection). If the initial state is
spatially homogeneous, a regime exists where homogeneity is
preserved~\cite{D00}, the temperature decays following the Haff's law
and any time-dependence is enslaved by the temperature. This regime,
known as Homogeneous Cooling State (HCS) lasts until instabilities
arise and homogeneity is broken.  In the HCS, it is useful to
introduce a new time-scale, defined by $d\tau=\omega(t) dt$ where
$\omega(t)$ is the time-dependent collision frequency and, at the same
time, rescaling density, velocity and temperature fluctuations by $n$
(which is constant), $\sqrt{2T(t)}$ and $T(t)$ respectively. With such
prescriptions one achieves a stationary representation of the HCS
whose Fourier-transformed linearized hydrodynamics has been
  studied for example in~\cite{BDKS98,NE00}. In this case, the
dynamical matrix reads

\begin{equation}A=
\begin{pmatrix}
  0                                &0                                    &0                      &0\\
-\gamma_0 g_2                &-(\gamma_0+\tilde{\ok} k^2)             &0                      &0\\
 0                                 &0                                    &\gamma_0-\tilde{\nl} k^2   &0\\
 0                                 &0                                    &0                      &\gamma_0-\tilde{\np} k^2
\end{pmatrix}\\+\mathcal{I}
\begin{pmatrix}
  0                                &0                                    &-k \lambda_0                    &0\\
  0                                &0                                    &-k \frac{2p_0}{dnT}\lambda_0    &0\\
 -k c^2 \lambda_0                &-k \frac{p_0}{2nT} \lambda_0      &0                           &0\\
 0                                 &0                                    &0                          &0
\end{pmatrix},
\label{Acool}
\end{equation}
and noise amplitudes are
\begin{eqnarray}
D_{nn}&=&0\\
D_{TT}&=& \frac{1}{2}\frac{4k^2 \ok}{n d \omega}\\
D_{\ul\ul}&=&\frac{1}{2}\frac{\nl k^2}{n\omega}\\
D_{\up\up}&=&\frac{1}{2}\frac{\np k^2}{n\omega}. \label{dvvcool}
\end{eqnarray}
Here $\tilde{\ok}=\ok/\omega$,
$\tilde{\nl}=\nl/\omega$ and $\tilde{\np}=\np/\omega$
are the rescaled transport coefficients, $\lambda_0$ is the mean free
path and $c^2=1/(2T)(\partial p/\partial n)_T$.  Note that the above
matrices are obtained by replacing $\gamma_b \to -\gamma_0\omega$ in
the matrices for the driven case (and of course applying the other
rescalings discussed above). Note also that the real part of some of
the eigenvalues of the hydrodynamical matrix become positive for
wave-vectors smaller than the cut-off value $k^*$: such an
instability, which marks the end of the HCS, can be prevented by
taking a system small enough.  Considering values of $k<k^*$, for the
HCS we find the following expression of the entropy production

\begin{equation} \label{nT3}
\langle\sigma_k\rangle=h''(k) \Re[\dn(k)\dot{T}^*(k)]
\end{equation}
with

\begin{equation}
h''(k)=-\frac{d \gamma_0 \omega  [g_2\nl+\frac{2p_0}{d n T} \lambda_0 (\ok+\nl)]}{2 \ok k^2 \nl}.
\end{equation}

\begin{figure}[!t]
\begin{center}
\includegraphics[width=0.6\columnwidth,clip=true]{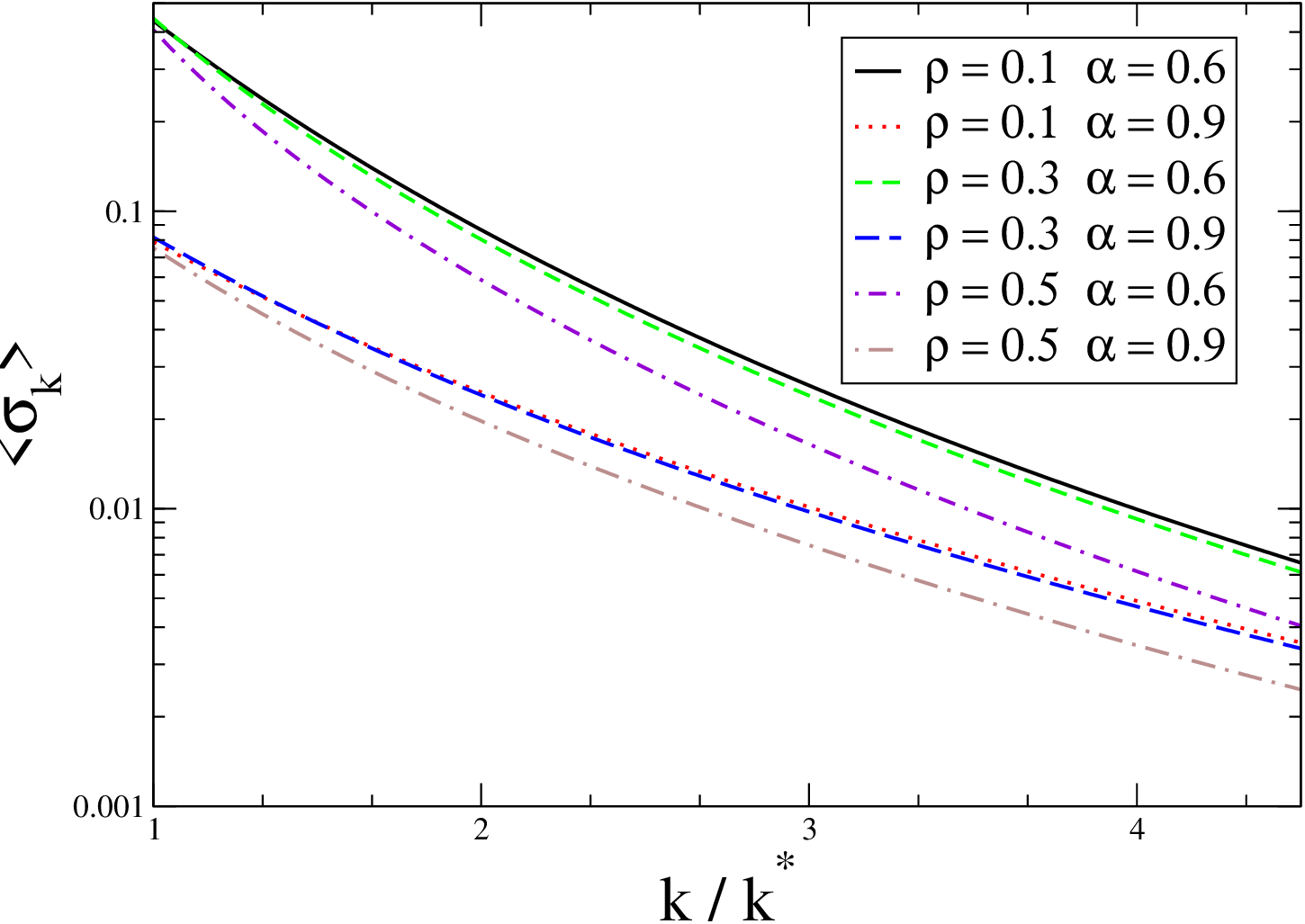}
\caption{ Average entropy production $\langle \sigma_k\rangle$ as a
  function of $k/k^*$, Eq.~\eqref{nT3}, for the homogeneous cooling state, defined by
  the Eqs.~(\ref{langgran}) and~(\ref{Acool}-\ref{dvvcool}). The
  different curves show the behavior of $\langle \sigma_k\rangle$ for
  several values of the parameters $\alpha$ and $\rho$, as reported in
  the legend. Notice that $\sigma_k$ reaches a finite value for
  $k=k^*$, where $k^*$ is the wave-vector of the instability in the
  system.}
\label{cool}
\end{center}
\end{figure}

In Fig.~\ref{cool} we show the behavior of the average entropy
production as a function of $k/k^*$, for different values of $\alpha$
and $\rho$.  Notice that, even approaching the value $k^*$, where
instabilities arise, the entropy production remains finite. Also in
this case, in the limit $k\to\infty$ we have the behavior
$\langle\sigma_k\rangle\sim k^{-4}$.

%%%%%%%%%%%%%%%%%%%%%%%%%%%%%%%%

\section{Conclusions}
\label{conclusions}

The coarse-grained dynamics of spatially extended systems can often be
described by a set of coupled Langevin equations. For this class of
stochastic systems, we have presented a general formula for the
fluctuating entropy production, showing that it can be put in a form
similar to that expected from macroscopic theories, such as linear
non-equilibrium thermodynamics. As an application, we have studied in
detail the fluctuating hydrodynamic equations describing the large
scales fluctuations of different kinds of granular fluids.  In
particular, we have singled out that, in homogeneously driven granular
fluids, the continuous flow of energy across the system produces some
correlations $\langle \delta n\dot{T}\rangle$, usually absent at
equilibrium. Namely, a density fluctuation in a certain point of space
couples to the variations in time of the local temperature in the
neighborhood. We identified the quantity $\delta n\dot{T}$ with the
out-of-equilibrium fluctuating current of such a system. Notice that,
due to the homogeneous driving mechanism, this current has no
direction in space (only in time), differently from many examples of
off-equilibrium systems.  Moreover, we have shown that the wave-vector
dependent entropy production, $\sigma_k$, is a quantity that provides
a measure of how much dissipative is the granular system when
variables are coarse-grained on a length-scale $\lambda \sim 1/k$.

%%%%%%%%%%%%%%%%%%%%%%%%%%%%%%%%
\begin{acknowledgments} 
  We thank E.~Trizac and D.~Villamaina for useful discussions.  The
  work is supported by the ``Granular-Chaos'' project, funded by the
  Italian MIUR under the FIRB-IDEAS grant number RBID08Z9JE.
\end{acknowledgments}

%%%%%%%%%%%%%%%%%%%%%%%%%%%%%%%%%%%%%%%%%%%%%%%%%%%%
\appendix

\appendix
\section{}
\label{app1}

In this Appendix we show that the entropy production vanishes
identically when equilibrium conditions are considered. In the
following sum over repeated indices is always meant. 

Our purpose here is to skecth the main idea, and this is best done -
for the sake of compactness and without losing in generality -
restricting to the case of all real fields. For such a case the entropy
production is given by

\begin{eqnarray}
\sigma'(t) &\equiv& \sigma(t)+ \ln p(\phi(0))-\ln p(\phi(t)) 
= D_{kk}^{-1}\int_0^tds B_{k,irr}(s)[\dot{\phi}_k(s)-B_{k,rev}(s)]+\ln p(\phi(0))-\ln p(\phi(t)) \nonumber \\
&=& D_{kk}^{-1}\int_0^tds (A^{irr})_{kj}\phi_j(s)\dot{\phi}_k(s)-
D_{kk}^{-1}\int_0^tds (A^{irr})_{kj}\phi_j(s)(A^{rev})_{kl}\phi_l(s) +\ln p(\phi(0))-\ln p(\phi(t)) \nonumber \\
&=& \frac{1}{2}D_{kk}^{-1}(A^{irr})_{kj}[\phi_j(t)\phi_k(t)-\phi_j(0)\phi_k(0)] 
-D_{kk}^{-1}\int_0^tds (A^{irr})_{kj}\phi_j(A^{rev})_{kl}\phi_l+\ln p(\phi(0))-\ln p(\phi(t)), \nonumber \\
\label{0.1}
\end{eqnarray}
where, following the definitions~(\ref{rev},\ref{irr}), 
$(A^{irr})_{ij}=1/2(A_{ij}+\epsilon_iA_{ij}\epsilon_j)$ and
$(A^{rev})_{ij}=1/2(A_{ij}-\epsilon_iA_{ij}\epsilon_j)$.  Using the
equilibrium distribution

\begin{equation}
p(\phi)\propto \exp[-\frac{1}{2}\phi_k(C^{-1})_{kl}\phi_l],
\label{0.2}
\end{equation}
together with the equilibrium relation

\begin{equation}
A^{irr}C=-D \Rightarrow D^{-1}A^{irr}=-C^{-1},
\label{0.3}
\end{equation}
one immediately finds that, in Eq.~(\ref{0.1}), the first term on the
rhs of the last line exactly cancels the quantity $\ln p(\phi(0))-\ln
p(\phi(t))$.  The remaining term can be recast in the form

\begin{equation}
D_{kk}^{-1}(A^{irr})_{kj}\phi_j(A^{rev})_{kl}\phi_l=-(C^{-1})_{kj}\phi_j(A^{rev})_{kl}\phi_l=
-(C^{-1})_{jk}\phi_j(A^{rev})_{kl}\phi_l =-\phi C^{-1}A^{rev}\phi,
\label{0.4}
\end{equation}
where the first equality follows from the relation~(\ref{0.3}), while
the property $C=C^{T}$ has been used in the second. Now, since at equilibrium

\begin{equation}
A^{rev}C+C (A^{rev})^T=0 \Rightarrow C^{-1}A^{rev}+(C^{-1}A^{rev})^T=0,
\label{0.5}
\end{equation}
one has that $C^{-1}A^{rev}$ is antisymmetric, namely 

\begin{equation}
\phi C^{-1}A^{rev}\phi=0.
\label{0.6}
\end{equation}
This implies $\sigma'=0$.

For the full complex case, the demonstration follows in a similar way.

\section{}
\label{app2}

In this appendix we present some complementary considerations in terms
of the Fokker-Planck equation associated with the Langevin
equations~(\ref{fw_eq}).  This allows us to discuss relations
involving the Onsager coefficients.

The Fokker-Planck equation obeyed by the probability density for a
linear process defined by Eqs.~\eqref{fw_eq} and~\eqref{lin_eq} reads
\begin{equation} \label{fp}
\frac{dp(\phi,t)}{dt}=-\sum_i \frac{\partial}{\partial \phi_i} [j_i(\phi,t) p(\phi,t)],
\end{equation}
with
\begin{equation}
j(\phi,t) = -A \phi - R f(\phi,t).
\end{equation}
Here $j$ is the probability current vector, $A$ the dynamical matrix,
\begin{equation}
f_i(\phi,t) = \frac{\partial \ln p(\phi,t)}{\partial \phi_i},
\end{equation}
and $R$ is {\em any} matrix which satisfies $R^s=D$ with $R^s$ its
symmetrized and $D$ the matrix of noise correlations (which is always
symmetric by definition). This arbitrariety on $R$ will be discussed
below. We anticipate that, at equilibrium, $R$ coincides with the
matrix of the Onsager coefficients $L$.

The stationary solution of Eq.~\eqref{fp}, $p(\phi)=\lim_{t\to
  \infty}p(\phi,t)$, satisfies
\begin{equation}
\ln [p(\phi)]=\textrm{const.}-\frac{1}{2} \phi G \phi^\dag,
\end{equation}
where $\phi^\dag$ is the adjoint of the vector $\phi$ (transpose and
conjugate) and $G=C^{-1}$ with $C_{ij}=\langle \phi_i\phi^*_j\rangle$
the self-adjoint matrix of covariances.  Let us also define the
quantity
\begin{equation}
f^0(\phi) = \lim_{t\to \infty}f(\phi,t) = -G\phi.
\end{equation}
By direct substitution one finds that $C$ must satisfy the
following equation:
\begin{equation}
-D = [AC]^s \equiv \frac{AC+CA^\dag}{2}.
\end{equation}
In general this means that
\begin{equation}
AC = R + Q
\end{equation}
with $Q$ an antihermitian matrix, i.e. $Q=-Q^\dag$. The relation with
the $D$ matrix is also ambiguous, i.e.  $AC = D + Q'$, with $Q'$
another (in principle different from $Q$) antihermitian matrix.  In
the stationary state one has, for the probability current:
\begin{equation}
j^0(\phi) = \lim_{t \to \infty} j(t)=-(R G)\phi-(Q G)\phi + (R G) \phi = - (Q G)\phi.
\end{equation}
Therefore, the probability current at any time may be rewritten as
\begin{equation} 
j(\phi,t) = - R[f(\phi,t)-f^0(\phi)] + j^0(\phi).
\label{currentonsager}
\end{equation}
The equilibrium condition is $j^0(\phi)=0$, which requires $Q = 0$,
leading to $R=AC$, which is the fluctuation-dissipation relation
(FDR). Moreover, at equilibrium, from the time
translation and time inversion invariance of the joint distribution
$p(\phi(t),t;\phi(0),0)=p(\overline{\phi(0)},t;\overline{\phi(t)},0)$,
there follow the relations

\begin{eqnarray}
C_{ij}&=&\epsilon_i\epsilon_j C_{ji}^* \\
L_{ij}&=&\epsilon_i\epsilon_j L_{ji}^*. \label{ons}
\end{eqnarray}
On the other hand, a non-equilibrium stationary state with $j^0 \neq
0$ can be observed for {\em open} systems where energy or matter is
exchanged with the boundaries/thermostats with a preferred direction
in time: this is the case of the driven granular fluids discussed in
Section~\ref{granular}.

Finally, note that in the Fokker-Planck equation, which is uniquely
determined by the linear Langevin equation, only $A$ and $D=R^s$ are
fixed (and consequently also $C$ or $G$). The matrix $R$ is determined
only in its hermitian part, while its antihermitian part is
undetermined. At equilibrium this indeterminacy disappears, because of
the FDR which fixes $R=AC$. Note also that the FDR does not imply $R$
(or $AC$) to be symmetric. Its symmetry properties are determined by
the parities under time-reversal~(\ref{ons}), as discovered by
Onsager~\cite{O31,O31b}: $R_{ij}=R_{ji}^*\epsilon_i\epsilon_j$.
Moreover, let us remark that the Onsager matrix is uniquely defined
only at equilibrium. Indeed one has at least two options to define the
Onsager matrix in the non-equilibrium case: 1) one can follow the
principle that the Onsager matrix is the ``proportionality factor''
between thermodynamic force $f-f^0$ and the current excess with
respect to $j^0$, so that Eq.~(\ref{currentonsager}) implies $L\equiv
R$; 2) one can follow the principle that the Onsager matrix is the
proportionality factor between the decay of fluctuations and the
stationary entropic force $X=-f^0$, i.e. $d\langle\phi\rangle/dt=-L\langle X\rangle$, and this
implies $L\equiv AC$. At equilibrium the two options coincide. Of
course those two principles are still arbitrary: Onsager matrix is,
indeed, {\em undefined} in non-equilibrium dynamics.

\end{document}